\documentclass[10pt,superscriptaddress,nofootinbib,notitlepage,tightenlines,twocolumn, pra]{revtex4-1}

\usepackage{amsthm,amsmath,amssymb}

\usepackage{mathrsfs}
\usepackage{amsthm}
\usepackage{amsmath}
\usepackage{amssymb}
\usepackage{graphicx}
\usepackage{epstopdf}
\usepackage{url}
\usepackage{float}
\usepackage{bm}
\usepackage{ifthen}
\usepackage[usenames,dvipsnames]{color}
\usepackage{mathrsfs}
\usepackage[colorlinks=true,citecolor=blue,urlcolor=black]{hyperref}
\usepackage{float}

\newcommand{\be}{\begin{equation}}
\newcommand{\ee}{\end{equation}}

\newcommand{\sket}[1]{{\ensuremath{\lvert#1\rangle}}}
\newcommand{\lket}[1]{{\ensuremath{\left\lvert#1\right\rangle}}}
\newcommand{\ket}[1]{\if@display\lket{#1}\else\sket{#1}\fi}

\newcommand{\sbra}[1]{{\ensuremath{\langle#1\rvert}}}
\newcommand{\lbra}[1]{{\ensuremath{\left\langle#1\right\rvert}}}
\newcommand{\bra}[1]{\if@display\lbra{#1}\else\sbra{#1}\fi}

\newcommand{\sbraket}[2]{{\ensuremath{\langle#1\rvert#2\rangle}}}
\newcommand{\lbraket}[2]{{\ensuremath{\left\langle#1\!\left\rvert\vphantom{#1}#2\right.\!\right\rangle}}}
\newcommand{\braket}[2]{\if@display\lbraket{#1}{#2}\else\sbraket{#1}{#2}\fi}

\newcommand{\sketbra}[2]{{\ensuremath{\lvert #1\rangle\!\langle #2\rvert}}}
\newcommand{\lketbra}[2]{{\ensuremath{\left\lvert #1\right\rangle\!\!\left\langle #2\right\rvert}}}
\newcommand{\ketbra}[2]{\if@display\lketbra{#1}{#2}\else\sketbra{#1}{#2}\fi}

\theoremstyle{plain}

\theoremstyle{definition}

\begin{document}

\title{Coherent one-way quantum conference key agreement based on twin field}
\author{Xiao-Yu Cao}
\affiliation{National Laboratory of Solid State Microstructures and School of Physics, Nanjing University, Nanjing 210093, China}
\author{Jie Gu}
\affiliation{National Laboratory of Solid State Microstructures and School of Physics, Nanjing University, Nanjing 210093, China}
\author{Yu-Shuo Lu}
\affiliation{National Laboratory of Solid State Microstructures and School of Physics, Nanjing University, Nanjing 210093, China}
\author{Hua-Lei Yin}\email{hlyin@nju.edu.cn}
\affiliation{National Laboratory of Solid State Microstructures and School of Physics, Nanjing University, Nanjing 210093, China}
\author{Zeng-Bing Chen}\email{zbchen@nju.edu.cn}
\affiliation{National Laboratory of Solid State Microstructures and School of Physics, Nanjing University, Nanjing 210093, China}

\begin{abstract}

Quantum conference key agreement (CKA) enables key sharing among multiple trusted users with information-theoretic security. Currently, the key rates of most quantum CKA protocols suffer from the limit of the total efficiency among quantum channels.
Inspired by the coherent one-way and twin-field quantum key distribution (QKD) protocols, we propose a quantum CKA protocol of three users. Exploiting coherent states with intensity 0 and $\mu$ to encode logic bits, our protocol can break the limit.
Additionally, the requirements of phase randomization and multiple intensity modulation are removed in our protocol, making its experimental demonstration simple.
\end{abstract}
\maketitle

\section{Introduction}
The establishment of quantum network is the ultimate goal of quantum communication, where QKD is the most mature subfield for applications.
QKD allows secret key sharing between two distant authorized participants with unconditional security~\cite{bennett1984proceedings,ekert1991quantum}.
But the number of participants in QKD is merely two.
In many other scenarios on quantum networks, such as web conference and online courses, there are far more than two users who need to share keys. Quantum CKA ~\cite{bose1998multiparticle,cabello2000multiparty,chen2007multi,matsumoto2007multiparty} gives the solution, which aims to distribute common secret keys among multiple parties. The advantages of quantum CKA over repeating QKD in quantum networks are that quantum CKA requires fewer resource qubits, transmits fewer classical bits and performs fewer rounds of error correction and privacy amplification steps, which have been illustrated by researchers~\cite{epping2017multi}.
After more than two decades of development, many quantum CKA protocols have been proposed~\cite{Fu:2015:Long,epping2017multi,proietti2020experimental,zhu2015w,wu2016continuous,ottaviani2019modular,chen2016biased,chen2017asymmetric,ribeiro2018fully,li2018quantum,pivoluska2018layered,jo2019semi,zhao2020phase,Yin2020breaking}, including measurement-device-independent~\cite{Fu:2015:Long} and device-independent~\cite{ribeiro2018fully} protocols. Details can be found in the review article~\cite{murta2020quantum}.

Nevertheless, the conference key rates in a majority of quantum CKA protocols are rigorously restricted by the total efficiency among quantum channels, which is an upper bound of the conference key rates~\cite{das2019universal}. Although some techniques may be utilized to break this upper bound, such as quantum repeaters~\cite{duan2001long}, adaptive measurement-device-independence~\cite{azuma2015all} and W states with single-photon interference~\cite{grasselli2019conference},
these approaches are difficult to be implemented practically.
Recently, a three-party quantum CKA~\cite{Yin2020breaking}, inspired by the twin-field QKD~\cite{lucamarini2018overcoming,Wang:2018:Twin,yin2019measurement}, is presented to overcome this limit in a practical way. However, this protocol~\cite{Yin2020breaking} requires to exploit the decoy-state method~\cite{wang2005beating,lo2005decoy}, including the active and high precision phase randomization and multiple intensity modulation, which increases the experimental complexity in quantum state preparation.

Here, we propose a simple scheme for three-party quantum CKA protocol to break this limit by combining the methods of coherent one-way QKD~\cite{stucki2005fast} and twin-field QKD~\cite{lucamarini2018overcoming}. The simulation results show that this protocol can be demonstrated over 450 km with available technology. Our protocol employs weak coherent states with intensity 0 or $\mu$ to encode logic bits. Therefore, phase randomization and pulse modulation with different intensities are circumvented in our protocol, which results in a much simpler experimental setup and decreased security risks caused by imperfect phase randomization.
The security proof of our protocol, surprisingly, can directly utilize the security analysis of coherent one-way QKD~\cite{stucki2005fast,branciard2008upper,korzh2015provably}. Our protocol promotes the practical process of quantum CKA and may have a good application prospect.

\section{Protocol description}\label{part2}

\begin{figure}[t!]
\centering
  \includegraphics[width=0.85\columnwidth]{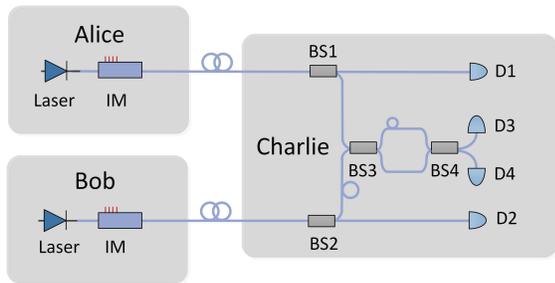}
  \caption{\textbf{The setup of our protocol.}\label{fig1}
  Weak coherent pulse sources (Laser); Intensity modulator (IM); beam splitter (BS1, BS2, BS3, BS4); detector (D1, D2, D3, D4). For Alice and Bob, they have same devices. While in Charlie's site, BS3 and BS4 form an asymmetric Mach-Zehnder interferometer.
 }
\end{figure}

As shown in figure~\ref{fig1}, there exist three parties in our protocol, named Alice, Bob and Charlie. Alice and Bob serve as senders while Charlie serves as a receiver. Since participants in our protocol encode logic bits with weak coherent states instead of single-photon states, our scheme features a simple experimental setup where senders only require a laser source and an intensity modulator to generate pulses.

Each sender prepares a pulse with intensity $\mu$ (0) for logic bit 1 (0).
They both send pulses with a period of $2T$.
Charlie measures the received pulses in time-interference bases and applies a passive-basis choice.
Bits are encoded in the time basis and coherence is checked in the interference basis.
For Charlie, we set that $\ket{0}_{\rm A}\ket{\alpha}_{\rm B}$ ($|\alpha|^2 = \mu$) means 0 and $\ket{\alpha}_{\rm A}\ket{0}_{\rm B}$ means 1. Charlie employs $D_1$ and $D_2$ to perform passive measurements in the time basis. Specifically, the bit value is 0 when only $D_2$ clicks in the time basis while the bit value is 1 when only $D_1$ clicks.
The distance between BS2-BS3 is longer than that between BS1-BS3 so as to delay Bob's pulses for a fixed time $T$ before entering the interferometer, which helps transform Alice’s and Bob’s pulse trains with the period of 2T into pulse trains with the period of T in two arms of the interferometer after BS3. The delay introduced in one arm of the interferometer is used to make the neighboring pulses interfere with each other.
Coherence can be quantified through the visibility of the interference.
We denote the time slot when Charlie performs interferometric measurements as $i$, with $i\in \{T, 2T, 3T, ..., 2NT\}$, where $N$ is the is the total number of pulses sent by Alice (Bob). And we set $k$ is an integer ranging from 1 to N. Visibility can be calculated by~\cite{stucki2005fast}:
\begin{equation}\label{equation1}
V=\frac{P(D_t)-P(D_f)}{P(D_t)+P(D_f)},
\end{equation}
where $P(D_t)$ is the probability that detector $D_3$ clicks when $i=(2k+1)T$ or $D_4$ clicks at $i=2kT$ and $P(D_f)$ is the probability that detector $D_4$ clicks at $(2k+1)T$ or $D_3$ clicks at $2kT$. The definition of visibility is similar to that in coherent one-way QKD and we relate the two schemes in Section~\ref{s3}. If the coherence of the signals is perfectly preserved, $P(D_f) = 0$ and $V=1$.

The whole process of our protocol is stated as follows:

\emph{1.~Preparation}. Alice (Bob) randomly prepares weak coherent pulses $\ket{\alpha}$ with probability $t$, and $\ket{0}$ with probability $1-t$.
Bob flips his logic bits after sending his pulses. They send the optical pulses to Charlie through insecure quantum channels.

\emph{2.~Measurement}.
Charlie measures the received pulses in time-interference bases.
Alice's (Bob's) pulse is split into two sub-pulses by BS1 (BS2). For Alice (Bob), one of two sub-pulses is sent to D1 (D2) for measurements in the time basis and the other two bunches of pulses are used for interferometric measurements.
Charlie records when and which detector clicks.
If different detectors click at the same time in the time basis, Charlie randomly chooses a bit value.
If detectors click at the same time in different bases, Charlie discards relevant data.

\emph{3.~Reconciliation}. Charlie announces the moment when detectors click and the corresponding basis information.
Alice and Bob publicly announce the intensity information only when detectors in interference basis click. Participants calculate visibility when there exist two neighboring $\ket{\alpha}$, i.e. $\ket{\alpha}_{b_k}\ket{\alpha}_{a_k}$ or $\ket{\alpha}_{a_{k+1}}\ket{\alpha}_{b_{k}}$.

\emph{4.~Parameter estimation}. Participants disclose a part of bit values from time basis to estimate the bit error rate and the gain of states, $\ket{0}\ket{\alpha}$ and $\ket{\alpha}\ket{0}$, which are used to encode logic bits. They evaluate the information Eve gets by calculating the visibility.

\emph{5.~Key distillation}. They extract the common conference key after classical error correction and privacy amplification.

\section{Security analysis}\label{s3}

\begin{figure*}[t!]
  \centering
  \includegraphics[width=1.5\columnwidth]{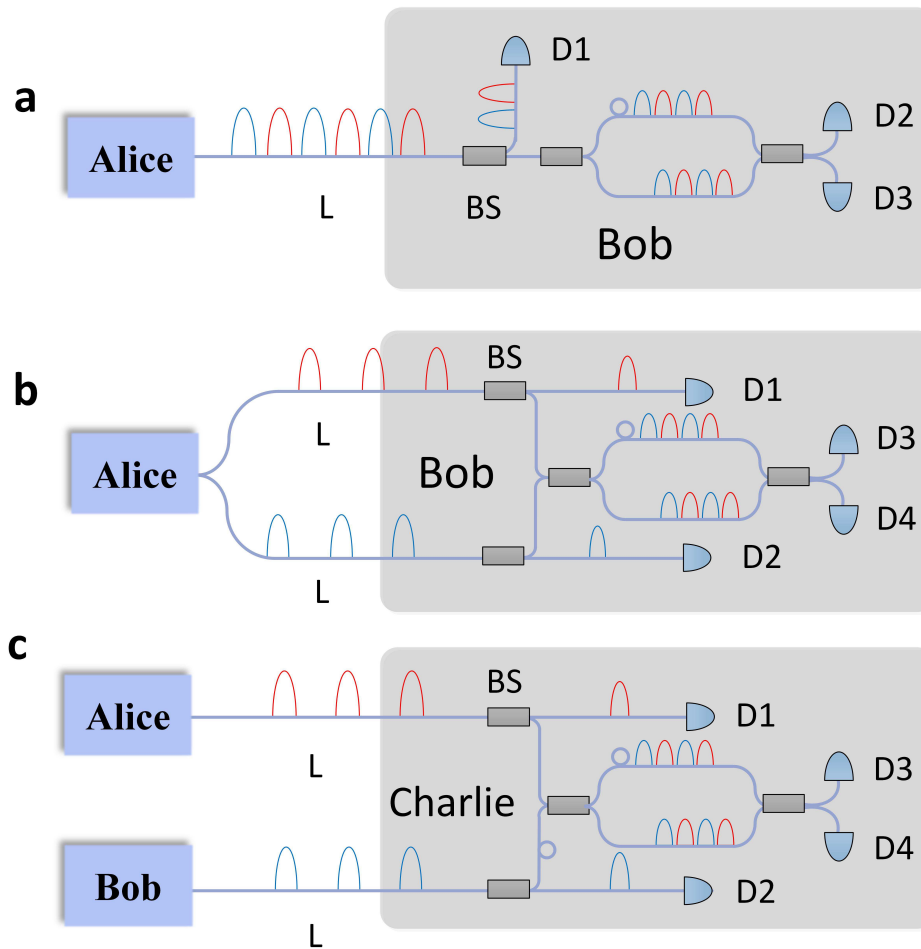}
  \caption{\textbf{The equivalence between coherent one-way QKD and our quantum CKA.}\label{fig2} (\textbf{a}). Typical coherent one-way QKD. Alice sends pulses and Bob detects them. The primary pulse is sent through an insecure quantum channel of length L and split by BS. One sub-pulse is used for encoding bits. Detector D1 measures the arrival time of pulses and determines the bit values by detection results.
  The other sub-pulse is used as the input of an asymmetric Mach–Zehnder interferometer.
  (\textbf{b}). Unfolded QKD setup. The common path of length L in \textbf{a} is split into two separate paths with length L. Red pulses and blue pulses travel on their own path, where blue ones are one-pulse time later than red ones. And they are split by different BSs. One sub-pulse of blue (red) pulse is detected by detectors D1 (D2).  The other two sub-pulses are the input of an asymmetric Mach–Zehnder interferometer.
  (\textbf{c}). Setup in this work. We use two light sources to send pulses independently. They send pulses at the same time. But the blue pulses are delayed for one-pulse time before interferometric measurements. }
\end{figure*}

We discuss the security of our protocol in this section. As depicted in figure~\ref{fig2}, we reduce our scheme to previous coherent one-way QKD~\cite{stucki2005fast}.
We first present the typical coherent one-way QKD set-up in figure~\ref{fig2}(a).
Alice sends pulses and Bob detects them. For both of them, logic bits are encoded with two-pulse sequences.
As an intermediate step towards our scheme, the common path of coherent one-way QKD
is unfolded in figure~\ref{fig2}(b). In this case, the two pulses travel on separate
channels and are encoded together. The rule of encoding logic bits is similar to above. Blue pulses are  one-pulse time later than red ones. After BS3, red and blue pulses with the period of 2T are transformed into pulse trains with the period of T in two arms of the interferometer, respectively. On short arm of the interferometer, the blue pulses have the opposite phase with other pulses, which is the only difference between figure~\ref{fig2}(a) and figure~\ref{fig2}(b) in the interference basis.
Therefore, we calculate visibility according to the parity of pulse's time slot which can be seen in equation~(\ref{equation1}).
The two schemes are equivalent from a security perspective.
In figure~\ref{fig2}(c), we present the quantum CKA scheme.
The detecting section have been outsourced to Charlie and the users’ stations have been separated.
Alice and Bob send pulses at the same time and Bob's (blue) pulses are postponed before entering the interferometer, which makes no difference in the measurements. Therefore, attacks on two different protocols have same influence. Thus we can take advantage of the security analysis of the existing protocol which is secure against a large class of collective attacks ~\cite{branciard2008upper,korzh2015provably}.

\section{Key rates}\label{part4}

By using the result of coherent one-way QKD (see also in~\ref{appA}),  the asymptotic conference key rate of our protocol can be written as
\begin{equation}
\begin{aligned}
R=&t(1-t)(Q_{0\alpha}+Q_{\alpha0})\Bigg[1-E_{\rm T}-(1-E_{\rm T})\\&h\left(\frac{1+\zeta(\mu,V)}{2}\right)\Bigg]
-Q_{\mu}fh(E_{\mu}),
\end{aligned}
\end{equation}
where Q$_{{\rm xy}}$ is the gain for the case of Alice choosing intensity x and Bob choosing intensity y, with x, y $\in \{0,\mu\}$.
$E_{\rm T}$ is the error rate under time basis and $h(x)$ is the binary Shannon entropy. $V$ is visibility in the interference basis.
$Q_{\mu}fh(E_{\mu})$ is the leaked information during classical error correction.
Here, $Q_{\mu}$ and $f$ are the overall gain under time basis and the error correction efficiency, respectively.
$E_{\mu}$ is the max error rate between X (X $\in \{ {\rm Alice, Bob} \}$) and Charlie.
$\zeta(\mu, V)=(2V-1)e^{-\mu}-2\sqrt{(1-e^{-2\mu})V(1-V)}$ from~\cite{korzh2015provably}.
More simulation details can be seen in~\ref{appB}.

\begin{table}[h]
\centering
\caption{\label{table}Simulation parameters~\cite{chen2020sending}. $\eta_{\rm d}$ and $p_{\rm d}$ are the detector efficiency and dark count rate. $\alpha$ is the attenuation coefficient of the ultralow-loss fiber. $f$ is the error correction efficiency and e$_{\rm d}$ is the misalignment rate of the time basis.}
\begin{tabular}{ccccc}
\hline
\hline
$\eta_{\rm d}$& ~~~~~~$p_{\rm d}$ &~~~~~~$\alpha$ & ~~~~~~$f$     &~~~~~~$e_{\rm d}$\\\hline
 $56\%$  &~~~~~~$10^{-8}$ &~~~~~~0.167&~~~~~~1.1& ~~~~~~$0.001$\\
\hline
\hline
\end{tabular}
\end{table}

\begin{figure}[t!]
  \centering
  \includegraphics[width=0.85\columnwidth]{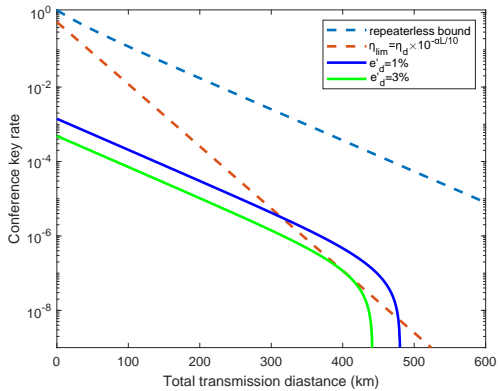}
  \caption{\textbf{Conference key rates using the experimental parameters in Table~\ref{table}. }\label{fig3} We define the misalignment rate under time basis as $e_{\rm d}=0.001$ and the misalignment rate under interference basis as
  $e'_{\rm d}$. We set $e'_{\rm d}=1\%$, and $3\%$ and compare their key rates to total efficiency among quantum channels
 and repeaterless bound.}
\end{figure}

Here, we assume that the sources and channels are all symmetric. The distances between Alice-Charlie and Bob-Charlie are L/2. Besides, the detection efficiencies and dark count rates of Charlie’s detectors are the same. The total efficiency of quantum channel is $\prod \limits_{i=1}^n\eta_i$, where $\eta_i$ is the efficiency of one quantum channel. The repeaterless bound between two users is PLOB bound and its form is $-\log_2(1-\eta)$~\cite{pirandola2017fundamental}, where $\eta$ is the channel efficiency between two users. Denote $\eta_{\rm lim}=\eta_{\rm d}\times 10^{-\alpha L/10}$ as the total efficiency among quantum channels and $\eta_{\rm rpl}=-\log_2(1-\eta_{\rm d}\times 10^{-\alpha L/20})$ as repeaterless bound~\cite{pirandola2020general,takeoka2019multipartite}, which is a generalized form of the two-user repeaterless bound~\cite{takeoka2014fundamental,pirandola2017fundamental}. We utilize the practical parameters which are presented in table~\ref{table} and optimize the conference key rate over the free parameters t and $\mu$.
By applying above-mentioned steps, the performance of our protocol is shown in figure~\ref{fig3}. Conference key rates in our protocol can break the limit of the total efficiency among quantum channels when the misalignment rate of interference basis $e'_{\rm d}$ is lower than $3\%$. Under this circumstance, the transmission distance of our protocol reaches over 450 km theoretically.

\section{Conclusion}

In summary, by employing coherent states, we have proposed a practical quantum CKA protocol that allows three parties to share secure conference keys.
Since Charlie encodes logic bits with a nonempty coherent state and a vacuum state, the key feature of our scheme is removing the requirement of coincidence detection, which is similar to the key idea of twin-field QKD.
Scaling as O($\sqrt{\eta}$) rather than O($\eta$),
the conference key rate of our practical protocol can surpass the limit of the overall efficiency among quantum channels when the misalignment rate of the interference basis is lower than $3\%$. The key rates of existing quantum CKA protocols are lower than that of repeating QKD and our protocol makes a contribution to narrowing the gap. Our protocol can be theoretically demonstrated over 450 km, whose performance is significantly beyond that in most previous works. Furthermore, encoding logic bits with coherent pulses, without the process of phase randomization and multiple intensity modulation, decreases the experimental difficulty and avoids the possible attacks caused by imperfect phase randomization.
Besides, a strong reference pulse for synchronization and the phase stabilization are also required in our protocol.

We believe that our practical quantum CKA protocol has wide application prospect and can be widely implemented in the approaching large-scale quantum network.
But note that our scheme still fails to beat the repeaterless bound and extending our scheme to N (${\rm N}>3$) parties is a nontrivial work. Unlike the twin-field QKD, our protocol is not a measurement-device-independent scheme. We hope evolving technologies can improve our protocol's performance.

\section{Acknowledgments}
We gratefully acknowledge support from the National Natural Science Foundation of China (under Grant No. 61801420); the Key Research and Development Program of Guangdong Province (under Grant No. 2020B0303040001); the Fundamental Research Funds for the Central Universities.

\appendix
\section{Coherent one-way QKD}\label{appA}

Coherent one-way QKD is one of the most practical protocols in the realm of quantum cryptography. In this section, we present a brief introduction to coherent one-way QKD~\cite{stucki2005fast}.

\begin{figure}[t!]
  \centering
  \includegraphics[width=0.85\columnwidth]{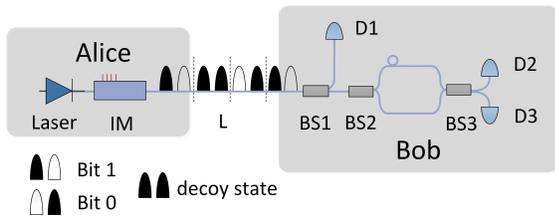}
  \caption{\textbf{The setup of coherent one-way QKD.}\label{fig5}  The pulses propagate to Bob are split at a beams splitter.
  The pulses transmitted to $D1$ are used to establish the raw key and the others going to the interferometer are used to check coherence.
  }
\end{figure}

There are 2 participants, Alice and Bob, in this scheme.
Alice sends pulses and Bob detects them.
As shown in figure~\ref{fig5}, Alice consists of an attenuated laser source followed by an intensity modulator and Bob owns beam splitters and detectors.
Alice prepares either a pulse of mean photon number $\mu$ ($\mu = |\alpha|^2$) or a vacuum pulse. The logic bits sent are encoded in the two-pulse sequence consisting of a nonempty and an empty pulse.
Bits 0 and 1 are encoded with $\ket{\alpha_0} := \ket{0} \ket{\alpha}$ and $\ket{\alpha_1} := \ket{\alpha} \ket{0}$.
Bob recovers the bit value simply by measuring the arrival time of the laser pulse.
To detect attacks on $\ket{\alpha_0}$ and $\ket{\alpha_1}$, Alice randomly sends a decoy state, $\ket{\alpha_t}:=\ket{\alpha}\ket{\alpha}$, to check for phase coherence between any two successive laser pulses. BS2 and BS3 form an asymmetric Mach–Zehnder interferometer.
The coherence of both decoy and 1-0 bit sequences can be checked with a single interferometer.

Detailed protocol is summarized as follows:

\emph{1.~Preparation}. Alice sends bit 0 or 1 with probability $\frac{1-t}{2}$ and the decoy sequence with probability $t$.

\emph{2.~Measurement}. Bob uses $D1$ to establish the raw key and the other detectors for interfermeotric measurements.

\emph{3.~Reconciliation}. When sufficient data is accumulated, Bob reveals for when detector clicked and the corresponding detector. Alice tells Bob which bits he has to remove from his raw key, since they come from detections of decoy sequences.

\emph{4.~Parameter estimation}. Alice evaluates the information Eve gets by analyzing the time and basis of clicks and calculating the visibility.

\emph{5.~Key distillation}. Alice and Bob run an error correction and a privacy amplification, and then extract the common conference key.

In~\cite{branciard2008upper}, security of coherent one-way QKD against collective attacks has been considered and an upper bound on the secret key rate for this protocol has been derived.
From~\cite{branciard2008upper,korzh2015provably}, we can know when comes to infinite key length, the key rate of coherent one-way QKD is
\begin{equation}
R = R_s\left[1-Q-(1-Q)h\left(\frac{1+\zeta(\mu, V)}{2}\right)\right]-leak_{\rm{EC}},
\end{equation}
where $R_s$ is the gain of pulses that can be used for encoding bits and Q is the quantum bit error rate.
$\zeta(\mu, V)=(2V-1)e^{-\mu}-2\sqrt{(1-e^{-2\mu})V(1-V)}$ and $leak_{{\rm EC}}$ means the leaked information during the process of error correction.

\section{Simulation details}\label{appB}

The asymptotic conference key rate of the ideal protocol is
\begin{equation}\label{ea2}
\begin{aligned}
R=&t(1-t)(Q_{0\alpha}+Q_{\alpha 0})\Bigg[1-E_{\rm T}-(1-E_{\rm T})\\
&h\left(\frac{1+\zeta(\mu, V)}{2}\right)\Bigg]
-Q_{\mu}fh(E_{\mu}),
\end{aligned}
\end{equation}
where Q$_{xy}$ is the gain for the case of Alice choosing intensity x and Bob choosing intensity y, with x, y $\in \{0,\mu\}$.
$E_{\rm T}$ is the error rate under time basis and $h(x)=-x\log_2x-(1-x)\log_2(1-x)$ is the binary Shannon entropy. $V$ is visibility in the interference basis. $Q_{\mu}$ and $f$ are the overall gain in the time basis and the error correction efficiency, respectively.
$E_{\mu}$ is the max error rate between X (X $\in \{ {\rm Alice, Bob} \}$) and Charlie.
$\zeta(\mu, V)$ is the same as that in coherent one-way QKD.

In the following part, we present how to obtain the values of parameters used in equation~(\ref{ea2}).
For coherent states, we take $\ket{k_a}_{\rm A}\ket{k_b}_{\rm B}$ as an example. When sending $\ket{k_a}_{\rm A}\ket{k_b}_{\rm B}$, the gain can be given by
\begin{equation}
Q_{k_a k_b} = 1 - (1 - 2p_{\rm d})e^{-(|k_a|^2+|k_b|^2)\eta}.
\end{equation}
$E_{k_a k_b}$ is the error rate when sending corresponding pulses and mainly owing to the misalignment
rate of basis, denoted as $e_{\rm d}$.
Besides, imperfect detectors also contributes to error rate and the error rate of this kind of errors is $e_0=\frac{1}{2}$.
It can be given by
\begin{equation}
E_{k_a k_b} =e_d + \frac{2p_{\rm d}(e_0 - e_{\rm d})e^{-(|k_a|^2+|k_b|^2) \eta}}{Q_{k_a k_b}}.
\end{equation}
$E_{\rm T}$ is the error rate under time basis,
which originates from $\ket{0}_{\rm A}\ket{\alpha}_{\rm B}$ and $\ket{\alpha}_{\rm A}\ket{0}_{\rm B}$,
while $E_{\rm V}$ means the error rate under interference basis,
which originates from $\ket{\alpha}_{\rm A}\ket{\alpha}_{\rm B}$.
By $E_{\rm T}(Q_{\alpha0}+Q_{0\alpha}) = E_{0\alpha}Q_{0\alpha}+E_{\alpha0}Q_{\alpha0}$, we can get $E_{\rm T}$,
The process of calculating $E_{\rm V}$ is similar to above.
But the pulses measured in interference basis experience more channel loss and we assume the overall transmission and detection efficiency under interference basis $\eta_{\rm V}=\eta/10$.
By $E_{\rm V} = \frac{1-V}{2}$, we can get the value of visibility.

The bit error rate between Alice-Bob is almost twice as that between Alice-Charlie or Bob-Charlie. Therefore, we set Charlie's raw key as reference key.
Since Alice and Bob are symmetric, we merely need to work out the error rate of Alice-Charlie.
For both Alice-Charlie and Bob-Charlie, as long as Alice and Bob send the same state and Charlie detects it, it will result in an inevitable error.
Thus, either of them has half the errors.
$\ket{\alpha}_{\rm A}\ket{0}_{\rm B}$ and $\ket{0}_{\rm A}\ket{\alpha}_{\rm B}$ also contribute to the error rate. In fact, their error rates here equal to $E_{\alpha0}$
and $E_{0\alpha}$. Obviously, $E_{0\alpha} = E_{\alpha 0}$ and $Q_{0\alpha} = Q_{\alpha 0}$, so
\begin{equation}
E_\mu = \frac{\frac{1}{2}\left((1-t)^2Q_{00}+t^2Q_{\alpha \alpha}\right)+2t(1-t)E_{0\alpha}Q_{0\alpha}}{Q_{\mu}}.
\end{equation}

For free parameters $\mu$ and $t$, we utilize Genetic Algorithm to find their optimal value.
Utilizing the practical experimental parameters in~\cite{chen2020sending} , we can get the relationship between the conference key rate and transmission distance.




\end{document}